\documentclass[aps,prl,twocolumn,superscriptaddress,showpacs]{revtex4}
\usepackage{epsfig}
\usepackage{color}
\usepackage{url}
\begin{document}
\preprint{\vbox{ \hbox{Belle Preprint 2006-23}
		 \hbox{KEK Preprint 2006-24}
}}
\title{
\quad\\[0.5cm]
\boldmath
Improved Measurements of Branching Fractions and $CP$ Partial Rate Asymmetries for $B \to \omega K$ and $B \to \omega \pi$}
\affiliation{Budker Institute of Nuclear Physics, Novosibirsk}
\affiliation{Chiba University, Chiba}
\affiliation{Chonnam National University, Kwangju}
\affiliation{University of Cincinnati, Cincinnati, Ohio 45221}
\affiliation{The Graduate University for Advanced Studies, Hayama, Japan} 
\affiliation{Gyeongsang National University, Chinju}
\affiliation{University of Hawaii, Honolulu, Hawaii 96822}
\affiliation{High Energy Accelerator Research Organization (KEK), Tsukuba}
\affiliation{Hiroshima Institute of Technology, Hiroshima}
\affiliation{University of Illinois at Urbana-Champaign, Urbana, Illinois 61801}
\affiliation{Institute of High Energy Physics, Vienna}
\affiliation{Institute of High Energy Physics, Protvino}
\affiliation{Institute for Theoretical and Experimental Physics, Moscow}
\affiliation{J. Stefan Institute, Ljubljana}
\affiliation{Kanagawa University, Yokohama}
\affiliation{Korea University, Seoul}
\affiliation{Kyungpook National University, Taegu}
\affiliation{Swiss Federal Institute of Technology of Lausanne, EPFL, Lausanne}
\affiliation{University of Ljubljana, Ljubljana}
\affiliation{University of Maribor, Maribor}
\affiliation{University of Melbourne, Victoria}
\affiliation{Nagoya University, Nagoya}
\affiliation{Nara Women's University, Nara}
\affiliation{National Central University, Chung-li}
\affiliation{National United University, Miao Li}
\affiliation{Department of Physics, National Taiwan University, Taipei}
\affiliation{H. Niewodniczanski Institute of Nuclear Physics, Krakow}
\affiliation{Nippon Dental University, Niigata}
\affiliation{Niigata University, Niigata}
\affiliation{University of Nova Gorica, Nova Gorica}
\affiliation{Osaka City University, Osaka}
\affiliation{Osaka University, Osaka}
\affiliation{Panjab University, Chandigarh}
\affiliation{Peking University, Beijing}
\affiliation{Princeton University, Princeton, New Jersey 08544}
\affiliation{RIKEN BNL Research Center, Upton, New York 11973}
\affiliation{University of Science and Technology of China, Hefei}
\affiliation{Seoul National University, Seoul}
\affiliation{Sungkyunkwan University, Suwon}
\affiliation{University of Sydney, Sydney NSW}
\affiliation{Tata Institute of Fundamental Research, Bombay}
\affiliation{Toho University, Funabashi}
\affiliation{Tohoku Gakuin University, Tagajo}
\affiliation{Tohoku University, Sendai}
\affiliation{Department of Physics, University of Tokyo, Tokyo}
\affiliation{Tokyo Institute of Technology, Tokyo}
\affiliation{Tokyo Metropolitan University, Tokyo}
\affiliation{Tokyo University of Agriculture and Technology, Tokyo}
\affiliation{Virginia Polytechnic Institute and State University, Blacksburg, Virginia 24061}
\affiliation{Yonsei University, Seoul}
 \author{C.-M.~Jen}\affiliation{Department of Physics, National Taiwan University, Taipei} 
 \author{P.~Chang}\affiliation{Department of Physics, National Taiwan University, Taipei} 
 \author{K.~Abe}\affiliation{High Energy Accelerator Research Organization (KEK), Tsukuba} 
 \author{K.~Abe}\affiliation{Tohoku Gakuin University, Tagajo} 
 \author{I.~Adachi}\affiliation{High Energy Accelerator Research Organization (KEK), Tsukuba} 
 \author{H.~Aihara}\affiliation{Department of Physics, University of Tokyo, Tokyo} 
 \author{D.~Anipko}\affiliation{Budker Institute of Nuclear Physics, Novosibirsk} 
 \author{K.~Arinstein}\affiliation{Budker Institute of Nuclear Physics, Novosibirsk} 
 \author{V.~Aulchenko}\affiliation{Budker Institute of Nuclear Physics, Novosibirsk} 
 \author{A.~M.~Bakich}\affiliation{University of Sydney, Sydney NSW} 
% ...
\collaboration{The Belle Collaboration}

\begin{abstract}
We report improved measurements of $B$ to pseudoscalar-vector decays containing an $\omega$ meson in the final state. 
Our results are obtained from a data sample that contains $388\times 10^{6}$ $B\overline{B}$ pairs accumulated at the 
$\Upsilon(4S)$ resonance, with the Belle detector at the KEKB asymmetric-energy $e^+ e^-$ collider. We measure the 
following branching fractions: 
${\mathcal B}(B^{+} \to \omega K^{+}) = [8.1\pm0.6({\rm stat.})\pm{0.6}({\rm syst.})]\times 10^{-6}$, 
${\mathcal B}(B^{+} \to \omega \pi^{+}) = [6.9\pm{0.6}({\rm stat.})\pm{0.5}({\rm syst.})]\times 10^{-6}$, and
${\mathcal B}(B^{0} \to \omega K^{0}) = [4.4^{+0.8}_{-0.7}({\rm stat.})\pm{0.4}({\rm syst.})]\times 10^{-6}.$
The partial width ratio $\frac{\Gamma (B^{+}\to \omega K^{+})}{\Gamma (B^{0}\to \omega K^{0})}$ = $1.7\pm0.3({\rm stat.}) 
\pm0.1({\rm sys.})$. We also set the $90$\% confidence level upper limit ${\mathcal B}(B^{0} \to \omega \pi^{0}) < 2.0\times 10^{-6}.$
In addition, we obtain the partial rate asymmetries ${\mathcal A}_{CP} = 0.05^{+0.08}_{-0.07}(\rm {stat.}) \pm 0.01(\rm {syst.})$ 
for $B^{+} \to \omega K^{+}$, and ${\mathcal A}_{CP} = -0.02 \pm 0.09(\rm {stat.}) \pm 0.01(\rm {syst.})$ for $B^{+}\to \omega \pi^{+}$.
\end{abstract}
\maketitle
%
%
\pagestyle{empty}
Charmless hadronic $B$ decays provide a rich ground to understand the dynamics of $B$ meson decays and the origin of $CP$ violation.
Two-body $B$ decays with a vector meson and a pseudoscalar particle $h$ ($h$ is either a kaon or a pion) proceed through combinations of
color-allowed ($T_u$) and color-suppressed ($C_u$) $b \to u$ tree diagrams and $b \to s$
($P_s$) or $b \to d$ ($P_d$) penguin diagrams. For example, $B^{+} \to \omega K^{+}$ proceeds through $T_u$, $C_u$ and $P_s$, 
$B^{0} \to \omega K^{0}$ proceeds through $P_s$, $B^{+} \to \omega \pi^{+}$ proceeds through $T_u$, $C_u$ and $P_d$, and 
$B^{0} \to \omega \pi^{0}$ proceeds through $C_u$ and $P_d$. The QCD factorization (QCDF) approach suggests that the branching fractions of 
$B^{+} \to \omega h^{+}$ decays are in the range $10^{-6}$--$10^{-5}$~\cite{QCDF}. The predicted ${\mathcal B}(B^{0}\to\omega K^{0})$ and 
${\mathcal B}(B^{+}\to\omega K^{+})$ are enhanced after including the rescattering effect from final-state interactions~\cite{QCDF2}. The 
perturbative QCD (PQCD) approach, on the other hand, suggests that ${\mathcal B}(B^{+}\to\omega K^{+})=3.22\times 10^{-6}$, and 
${\mathcal B}(B^{0}\to\omega K^{0})=2.07\times 10^{-6}$, and that ${\mathcal B}(B^{+}\to\omega K^{+})=(10.6^{+10.4(+7.2)}_{-5.8(-4.4)})\times 10^{-6}$, 
and ${\mathcal B}(B^{0}\to\omega K^{0})=(9.8^{+8.6(+6.7)}_{-4.9(-4.3)})\times 10^{-6}$ after considering next-to-leading-order accuracy~\cite{PQCD}. 
Due to the lack of $T_u$ and $P_s$ diagrams, $B^{0} \to \omega \pi^{0}$ is expected to be small~\cite{QCDF, PQCD2}. Experimentally, clear signals 
have been observed for $B^{+} \to \omega K^{+}, B^{+} \to \omega \pi^{+}$, and $B^{0} \to \omega K^{0}$ with similar branching fractions~\cite{chwang, babar, babar2, babar3}. 
However, experimental measurements to date are not yet precise enough for a quantitative confirmation of the pattern predicted by QCDF or PQCD.\par   
In this paper, we report improved measurements of branching fractions and partial rate asymmetries for $B \to \omega h$ decays, where $h$ can 
be a kaon or pion. The partial rate asymmetry (${\mathcal A}_{CP}$) is measured for the charged $B$ decays and defined to be 
\begin{eqnarray}  
{\mathcal A}_{CP} \equiv \frac{\Gamma(B^{-} \to \omega h^{-})-\Gamma(B^{+} \to \omega h^{+})}{\Gamma(B^{-} \to \omega h^{-})+\Gamma(B^{+} \to \omega h^{+})}.
\end{eqnarray}
These measurements are based on a data sample of $388 \times 10^{6}$ $B\overline{B}$ pairs collected with the Belle detector at the KEKB~\cite{KEKB} 
asymmetric-energy $e^{+}e^{-}$ collider. They improve upon our previously published results~\cite{chwang} by a five-fold increase in statistics and 
supersede them.
The Belle detector is a large-solid-angle magnetic spectrometer that consists of a silicon vertex detector (SVD), a 50-layer 
central drift chamber (CDC), an array of aerogel threshold \v{C}erenkov counters (ACC), a barrel-like arrangement of time-of-flight scintillation 
counters (TOF), and an electromagnetic calorimeter (ECL) comprised of CsI(Tl) crystals located inside a superconducting solenoid coil that provides a 
$1.5\ \rm{T}$ magnetic field. An iron flux-return located outside of the coil is instrumented to detect $K_{L}^{0}$ mesons and to identify muons 
(KLM). The detector is described in detail elsewhere~\cite{Belle}. In August 2003, the three-layer SVD was replaced by a four-layer radiation 
tolerant device. The data sample for this analysis consists of $140$ $\rm{fb}^{-1}$ of data recorded with a three-layer 
SVD (Set I)~\cite{Belle} and $217$ $\rm{fb}^{-1}$ recorded with a four-layer SVD (Set II)~\cite{SVD2}.\par
Hadronic events are selected using criteria based on the charged track multiplicity and total visible energy, with an efficiency greater than 
$99$\% for generic $B\overline{B}$ events. All primary charged tracks must satisfy quality requirements based on their impact parameters relative 
to the run-dependent interaction point (IP). For tracks from the candidate $B$ mesons, their deviations from the IP position are required to be 
within $\pm 0.1\ \rm{cm}$ in the transverse direction and $\pm 3.0\ \rm{cm}$ in the longitudinal direction.\par
Charged particle identification is performed using a $K$-$\pi$ likelihood ratio, ${\mathcal R}_{K} = {\mathcal L}_{K}$/(${\mathcal L}_{\pi}+{\mathcal L}_{K}$), 
where ${\mathcal L}_{K}$ (${\mathcal L}_{\pi}$) is the likelihood for a charged particle to be a kaon (pion) based on information from the ACC, 
TOF and CDC. Charged tracks with ${\mathcal R}_{K} > 0.6$ are identified as kaons, and tracks with ${\mathcal R}_{K} < 0.4$ 
are identified as pions. With these criteria, kaons produced in $B^{+} \rightarrow \omega K^{+}$ decays are selected with an efficiency of $85$\%, 
while the corresponding rate of kaons that are misidentified as pions is $11$\%. On the other hand, pions from $B^{+} \to \omega \pi^{+}$ are 
selected with an efficiency of $90$\% and have a corresponding kaon misidentification rate of $9$\%.\par
Candidate $\pi^{0}$ mesons are reconstructed from pairs of photons with an invariant mass in the range of 
$118\ {\rm MeV}/c^{2}<$ $M_{\gamma\gamma}$ $< 150\ {\rm MeV}/c^{2}$ (within $\pm 2.5\sigma$ around the PDG value~\cite{PDG}). 
The cosine of the decay angle should satisfy $|\cos\theta_{\gamma}|<0.9$, where $\theta_{\gamma}$ is the angle between the photon decay axis 
and the negative of the laboratory frame direction in the $\pi^0$ rest frame. Candidate $K^{0}_{S}$ mesons are reconstructed using pairs of 
oppositely charged tracks that have an invariant mass in the range of $482\ {\rm MeV}/c^{2}<$ $M_{\pi^{+}\pi^{-}}$ $< 514\ {\rm MeV}/c^{2}$ 
(within $\pm 4\sigma$ around the PDG value~\cite{PDG}). The vertex of the $K^{0}_{S}$ candidate is required to be well reconstructed and 
displaced from the IP, and the $K^{0}_{S}$ momentum direction must be consistent with the $K^{0}_{S}$ flight direction. Candidate 
$\omega \to \pi^{+}\pi^{-}\pi^{0}$ decays are reconstructed from charged tracks with ${\mathcal R}_{K} < 0.9$ (this requirement has a $96$\% 
efficiency per track), and from $\pi^{0}$ candidates with $e^{+}e^{-}$ center-of-mass frame (CM) momenta greater than $0.35\ {\rm GeV}/c^{2}$. 
Candidate $\omega$ mesons are required to have invariant masses within $\pm 30\ {\rm MeV}/c^2$ ($\sim \pm 3\sigma$) of the PDG value~\cite{PDG}.\par
$B$ meson candidates are formed by combining an $\omega$ meson with either a kaon ($K^{+}$, $K^{0}_{S}$) or a pion ($\pi^{+}$, $\pi^{0}$).
Two kinematic variables are used to select $B$ candidates: the beam-energy constrained mass $M_{\rm bc}$ = 
$\sqrt{(E^{\rm CM}_{\rm beam})^2-|P^{\rm CM}_{B}|^2}$, and the energy difference $\Delta E = E^{\rm CM}_{B} - E^{\rm CM}_{\rm beam}$, where 
$E^{\rm CM}_{\rm beam}$ is the beam energy in the CM frame, and $P^{\rm CM}_{B}$, $E^{\rm CM}_{B}$ are the momentum and 
energy, respectively, of the $B$ candidate in the CM frame. Candidates with $M_{\rm bc} > 5.2\ {\rm GeV}/c^2$ and 
$|\Delta E|<0.25\ \rm{GeV}$ ($|\Delta E|<0.30\ \rm{GeV}$ for $B^{0} \to \omega \pi^{0}$) are selected for further analysis.\par
The dominant background arises from quark-antiquark continuum events ($e^{+}e^{-} \to q\overline{q}$, $q = u$,$d$,$s$,$c$). The continuum background 
is characterized by a jet-like structure, while the $B\overline{B}$ events have a more spherical distribution. The following event-shape variables 
calculated in the CM frame are employed to suppress the continuum. The thrust angle $\theta_{T}$ is defined as the angle between 
the thrust axis~\cite{thrust} of the $B$ candidate daughter particles and that of the rest of the particles in an event. Signal events are uniformly 
distributed in $\cos\theta_{T}$, while continuum events are sharply peaked near $\cos\theta_{T} = \pm 1.0$. Events with $|\cos\theta_{T}| < 0.9$ 
are selected.\par
A Fisher discriminant is formed by combining a set of modified Fox-Wolfram moments~\cite{SFW} with the variable $S_{\perp}$~\cite{sper}. This variable
is the scalar sum of the transverse momenta of particles outside a $45^{\circ}$ cone around the $B$ thrust axis, divided by the scalar sum of 
their momenta. Here, the transverse momenta are calculated with respect to the thrust axis, and we do not include daughters of the $B$ candidate.  
Further variables that have been found to separate signal from continuum background include: the cosine of the angle between the flight 
direction of the $B$ candidate and the beam direction ($\cos\theta_{B}$); the distance along the beam direction between the $B$ vertex and 
the vertex of the remaining particles in the event ($\Delta z$); and the cosine of the helicity angle, defined as the angle
between the direction opposite to the $B$ flight direction in the $\omega$ rest frame and the direction normal to the plane 
defined by the three daughter pions of the $\omega$. The probability density functions (PDFs) for these three variables and the Fisher discriminant 
are obtained using Monte Carlo (MC) simulation for signal events and sideband data ($M_{\rm bc} < 5.26\ {\rm GeV}/c^{2}$) for $q\overline{q}$ backgrounds. 
These variables are combined to form a likelihood ratio ${\mathcal R} = \frac{{\mathcal L}_{S}}{{\mathcal L}_{S}+{\mathcal L}_{q\overline{q}}}$, where 
${\mathcal L}_{S (q\overline{q})}$ is the product of signal ($q\overline{q}$) PDFs.\par
Additional background discrimination is provided by the quality of the $B$ flavor tagging of the accompanying $B$ meson. We use the standard 
Belle $B$ tagging package~\cite{TaggingNIM}, which gives two outputs: a discrete variable ($q$) indicating the $B$ flavor, and a dilution factor 
($r$) ranging from zero for no flavor information to unity for unambiguous flavor assignment. We divide the data into six $r$ regions. Continuum 
suppression is achieved by applying a mode-dependent requirement on ${\mathcal R}$ for events in each $r$ region based on the figure-of-merit:
${\mathcal N}_{s}/\sqrt{{\mathcal N}_{s}+{\mathcal N}_{q\overline{q}}}$, where ${\mathcal N}_{s}$ is the expected number of signal events 
estimated from simulation and our previously published branching fractions~\cite{chwang}, and ${\mathcal N}_{q\overline{q}}$ is the number of background 
events estimated from sideband data ($M_{\rm bc} < 5.26\ {\rm GeV}/c^{2}$). This ${\mathcal R}$ requirement retains $74$\%, $68$\%, $74$\%, 
and $57$\% of the signal while rejecting $91$\%, $94$\%, $91$\%, and $95$\% of the continuum background for the $\omega K^{+}$, 
$\omega \pi^{+}$, $\omega K^{0}$, and $\omega \pi^{0}$ modes, respectively.\par 
Simulation studies indicate small backgrounds from generic $b\to c$ transitions in the charged $B$ modes; they are found to be negligible for 
the neutral modes. Two additional backgrounds have also been considered: reflections of $B \to \omega \pi^{+}$ decays due to $\pi\ \to K$
misidentification, and feed-down from other charmless $B$ decays, predominantly $B\to \omega K^*(892)$ and $B\to \omega\rho(770)$. We include 
these three components in the fit used to extract the signal.\par
The signal yields and partial rate asymmetries are obtained using an extended unbinned maximum likelihood (ML) fit for $M_{\rm bc}$ and $\Delta E$. 
The likelihood is defined as 
\begin{eqnarray}
\mathcal{L} & = & e^{-\sum_{j}{\mathcal N}_{j}}
\times \prod_{i} \left [\sum_{j} {\mathcal N}_{j} \mathcal{P}_{j}(M_{{\rm bc}i}, \Delta E_{i})\right ],
\end{eqnarray}
where $i$ is the identifier of the $i$-th event, $P_{j}(M_{{\rm bc}i}, \Delta E_{i})$ is the PDF of $M_{\rm bc}$ and $\Delta E$, ${\mathcal N}_{j}$ 
is the number of events for category $j$, which corresponds to either signal, $q\overline{q}$ background, reflections due to $K$-$\pi$ misidentification, 
or $B\overline{B}$ background ($b\to c$ and charmless). There is no reflection component or $b\to c$ background for the neutral $B$ modes.\par 
The signal distribution in $M_{\rm bc}$ is parameterized by a Gaussian function centered near the mass of the $B$ meson, while the 
Crystal Ball line shape~\cite{crystal} is used to model the $\Delta E$ distribution. Both functions are calibrated with large 
control samples of $B \to {\overline D}{}^{0} \pi$, ${\overline D}{}^{0} \to K^{+}\pi^{-}\pi^{0}$, and $B \to \eta^{\prime} K$, 
$\eta^{\prime} \to \eta\pi^{+}\pi^{-}$, $\eta \to \gamma\gamma$ decays. The continuum PDF is the product of a first-order polynomial for
$\Delta E$ and an ARGUS function~\cite{argus} for $M_{\rm bc}$. All parameters of the continuum PDF are allowed to float in the fit except for 
the endpoint of the ARGUS function, which is determined using the $B \to {\overline D}{}^{0} \pi$ sample. Other background PDFs are modelled by 
a smoothed two-dimensional $M_{\rm bc}-\Delta E$ function obtained from MC simulation. The yields of the continuum and $B\overline{B}$ backgrounds 
are floated in the fit, except for the reflection component. The normalizations of the reflection components are fixed to expectations based
on the $B^{+} \to \omega K^{+}$ and $B^{+} \to \omega \pi^{+}$ branching fractions, as well as $K^{+} \leftrightarrow \pi^{+}$ fake rates. The 
reflection yields are first calculated according to the assumed $\omega K^+/\pi^+$ branching fractions and then re-estimated based on our measured 
results from the fit. After a few iterations, the reflection yields converge. For the charged $B$ modes, we perform independent fits to the 
$B^+$ and $B^-$ samples in order to extract ${\mathcal A}_{CP}$ according to Eq.~1.\par
Figure~1 shows the data and fit results for each mode. Table~\ref{prd06-2} lists the resulting event yields, efficiencies, 
corresponding branching fractions and ${\mathcal A}_{CP}$. These new results are consistent with those measured by BaBar~\cite{babar3}.
The branching fractions are computed as the sum of the yields divided by the corresponding efficiencies in each dataset and the total number 
of $B$ mesons. The reconstruction efficiency for each mode includes all intermediate branching fractions and is defined as 
the fraction of the signal yield remaining after all selection criteria, where the yield is determined by performing the unbinned maximum 
likelihood fit on a MC sample. Small correction factors are included to account for differences in the reconstruction efficiency between MC 
and data. The correction factor is obtained as a result of the relative $\pi^{0}$ reconstruction ($1.00$ for Set I and $0.96$ for Set II) 
using the inclusive $\overline{D}{}^{0} \to K^+\pi^-$ and ${\overline D}{}^{0} \to K^+\pi^-\pi^0$ samples. The corresponding systematic error 
is $4$\%. We determine the identification efficiency of charged kaons and pions by studying an inclusive $D^{*-} \to {\overline D}{}^{0}\pi^{-}$, 
$\overline{D}{}^{0} \to K^+\pi^-$ sample. This selection criterion leads to a correction factor within the range of $0.90$--$0.96$ and 
a corresponding systematic error of $1.7$\% for charged modes and $0.9$\% for neutral modes.\par 
\begin{table*}
\caption{Signal efficiency ($\epsilon$), signal yield (${\mathcal N}_{{\rm s}}$), significance including systematic error 
($\Sigma_{\rm {sig.}}$), branching fraction (${\mathcal B}$), the $90$\% C.L. upper limit (U.L.) and partial rate 
asymmetry (${\mathcal A}_{CP}$) for the $B^{+} \to \omega h^{+}$ decays. The first and second errors are statistical and 
systematic, respectively.}
\vspace{0.1cm}
\begin{tabular}{|l c c c c c c|}\hline\hline
\centering
Mode & $\epsilon$ (\%) & ${\mathcal N}_{s}$ & $\Sigma_{\rm {sig.}}$ & ${\mathcal B}$ ($10^{-6}$) & 
$\rm {U.L.}$ ($10^{-6}$) & ${\mathcal A}_{CP}$ \\\hline    
$\omega K^{+}$ & $8.27\pm0.17$ & $259.5^{+20.4}_{-19.4}$ & 19.5 & $8.1\pm0.6\pm{0.6}$ & $-$ & $0.05^{+0.08}_{-0.07} \pm0.01$ \\ [0.1cm] 
$\omega \pi^{+}$ & $8.43\pm0.17$ & $224.8^{+20.3}_{-19.3}$ & 17.1 & $6.9\pm{0.6}\pm{0.5}$ & $-$ & $-0.02 \pm0.09 \pm0.01$ \\ [0.1cm]
$\omega K^{0}$ & $2.50\pm0.09$ & $41.5^{+8.0}_{-7.0}$ & 9.3 & $4.4^{+0.8}_{-0.7}\pm{0.4}$ & $-$ & $-$ \\ [0.1cm]     
$\omega \pi^{0}$ & $3.80\pm0.12$ & $5.9^{+4.8}_{-4.1}$ & 1.5 & $0.5^{+0.4}_{-0.3}\pm{0.1}$ & $2.0$ & $-$ \\ [0.1cm]\hline\hline
\end{tabular}
\label{prd06-2}
\end{table*}
\begin{table}
\caption{systematic uncertainties for $\omega K$ and $\omega \pi$ (\%).}
\vspace{0.1cm}
\begin{tabular}{|l|c c c c|} \hline\hline
\centering
Mode\ & $\omega K^{+}$ & $\omega \pi^{+}$ & $\omega K^{0}$ & $\omega \pi^{0}$ \\\hline    
track reconstruction                                          & 3.0 & 3.0 & 4.0 & 2.0 \\ [0.1cm]
$\mathcal R$ requirement                                      & 2.5 & 2.6 & 2.5 & 3.3 \\ [0.1cm]
particle identification                                       & 1.7 & 1.7 & 0.9 & 0.9 \\ [0.1cm]
$\pi^0$ reconstruction                                        & 4.0 & 4.0 & 4.0 & 8.0 \\ [0.1cm]
$K^0_S$ reconstruction                                        & $-$ & $-$ & 4.9 & $-$ \\ [0.1cm]
$\omega$ mass window                                          & 3.0 & 3.0 & 3.0 & 3.0 \\ [0.1cm]		 	 
MC statistics                                                 & 1.1 & 1.2 & 1.8 & 2.4 \\ [0.1cm] 
signal PDF                                                    & $^{+0.1}_{-0.0}$ & $\le 0.1$ & $\pm0.5$ & $^{+1.2}_{-0.6}$ \\ [0.1cm]
$\omega K^+/\omega \pi^+$ feed-across background              & $^{+0.3}_{-0.1}$ & $\pm0.3$ & $-$ & $-$ \\ [0.1cm] 
$N_{B\bar B}$                                                 & 1.3 & 1.3 & 1.3 & 1.3 \\ [0.1cm]\hline 
Total                                                         & 6.8 & 6.8 & 8.8 & 9.9 \\ [0.1cm]\hline\hline
\end{tabular}
\label{syserr}
\end{table}
\begin{figure}
\centering
\includegraphics[height=405pt, width=0.52\textwidth]{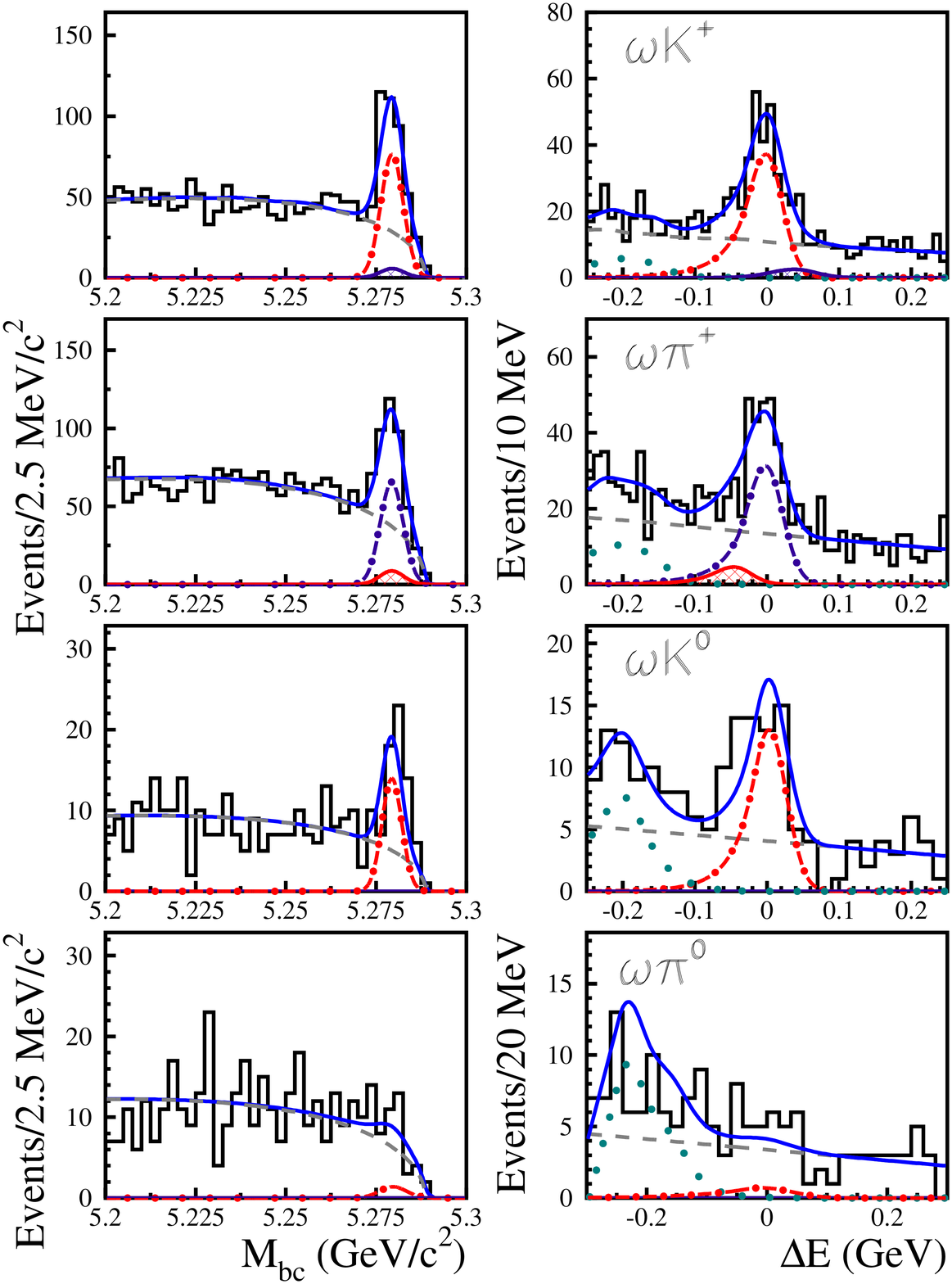}
\caption{Projections of fit results on $M_{\rm bc}$ (within the region of $|\Delta E|\leq0.1$\ GeV, 
$-0.15\ {\rm GeV}\leq\Delta E\leq0.1$\ GeV for $\omega \pi^{0}$) and $\Delta E$ (within the region of 
$5.27\ {\rm GeV}/c^2 <M_{\rm bc}<5.29\ {\rm GeV}/c^2$) for $\omega K^{+}$, $\omega \pi^{+}$, $\omega K^{0}$, 
and $\omega \pi^{0}$. Open histograms are data, solid curves are the fit functions, solid-dotted lines represent the signals, 
dashed lines show the sum of $q\overline{q}$ continuum and $b \to c$, dotted lines are due to other charmless $B$ decays, 
and cross-hatched areas represent reflections due to $K$-$\pi$ misidentification.} 
\label{omegahfit-2}
\end{figure}
\begin{figure}
\centering
\includegraphics[height=380pt, width=0.48\textwidth]{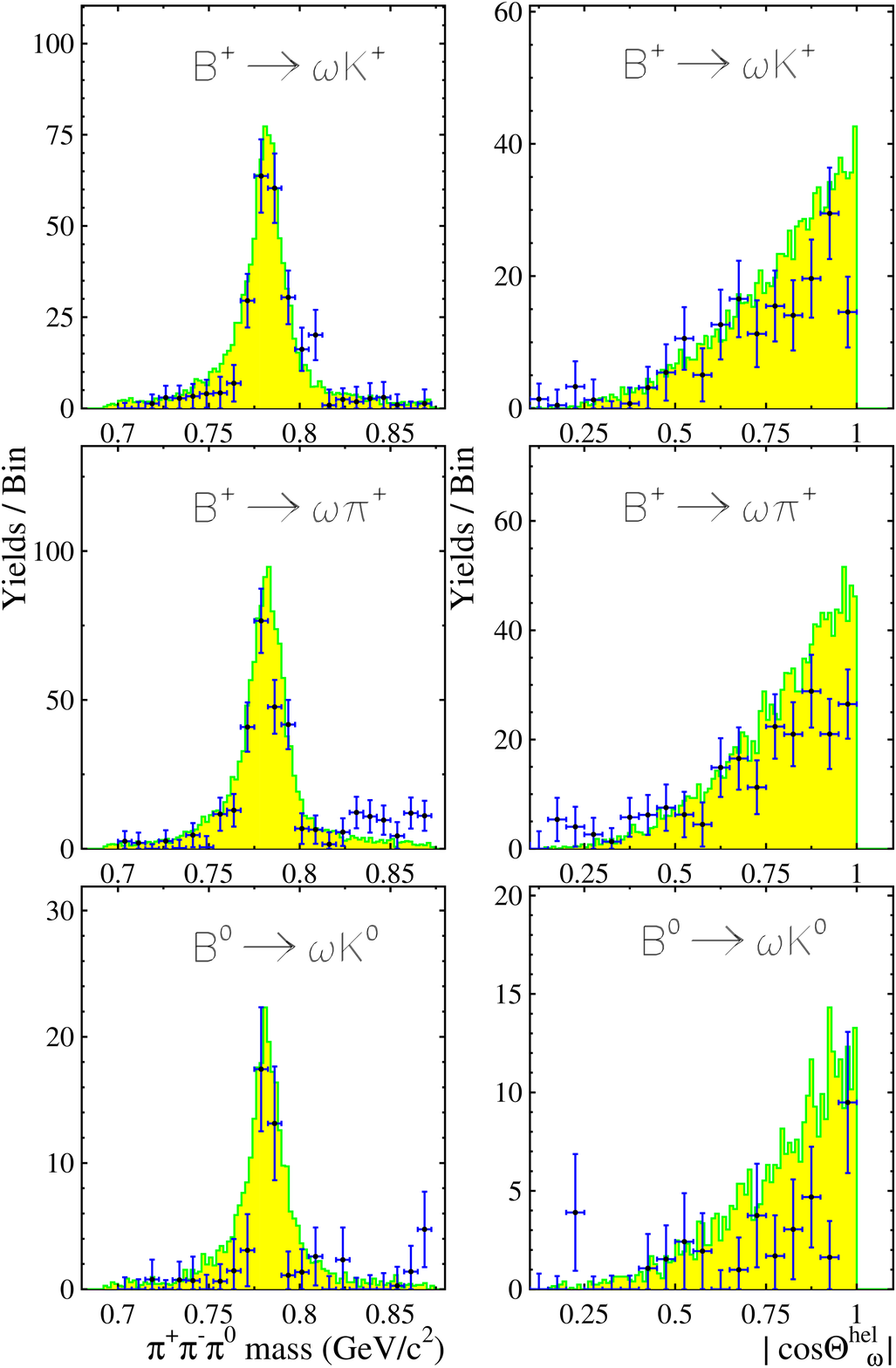}
\caption{Background-subtracted one-dimentional projections are obtained from data fit results in bins of fit yields in (left) 
the invariant mass ${\mathcal M}_{\omega \to \pi^+\pi^-\pi^0}$ and (right) the absolute value of the cosine of the helicity angle 
$|\cos\theta^{\omega}_{\rm hel}|$ for $\omega K^+$, $\omega \pi^+$, and $\omega K^0$ candidate events. Points are data, and 
histograms are from MC expectations for $B\to \omega h$.}
\label{crosscheck}
\end{figure}
We define the significance as $\sqrt{-2{\rm ln}({\mathcal L}_{0}/{\mathcal L}_{\rm max})}$, where ${\mathcal L}_{\rm max}$ is the maximum 
likelihood from the fit when the signal yields are floated, and ${\mathcal L}_{0}$ is obtained when the signal yields are set to zero with the 
remaining fit parameters still floating. To verify that signal candidates indeed arise from $B \to \omega K$ or $B \to \omega\pi$, we in turn 
change the criterion on ${\mathcal M}_{\omega\to\pi^+\pi^-\pi^0}$ or $|\cos\theta^{\omega}_{\rm hel}|$, and repeatedly perform ML fits for 
$M_{\rm bc}$ and $\Delta E$ to obtain yields in each bin of ${\mathcal M}_{\omega \to \pi^+\pi^-\pi^0}$ or $|\cos\theta^{\omega}_{\rm hel}|$ 
shown in Figure~2, where $\theta^{\omega}_{\rm hel}$ denotes the angle between the $B$ flight direction and the normal to the decay plane in 
the $\omega$ rest frame. The $90$\% C.L. upper limit for ${\mathcal B}(B^{0} \to \omega \pi^{0})$ is calculated using the procedure of Ref.~\cite{POLE}, 
which is based on the Feldman-Cousins method~\cite{UL}. This procedure accounts for systematic uncertainties (discussed below). The estimated 
background (observed count) is $57$ ($65$) events inside the $M_{\rm bc}-\Delta E$ signal box. \par 
The main systematic uncertainties for the branching fraction measurements are listed in Table~\ref{syserr}. These are evaluated as follows: 
charged tracking efficiency ($1.0$\% per track) from partially reconstructed $D^{*+} \to D^{0}\pi^{+}$, $D^0 \to K^{-}\pi^{+}$; $K^{0}_{S}$ 
reconstruction ($4.9$\%) from $D^{+} \to K^{0}_{S}\pi^{+}$, $D^{+} \to K^{-}\pi^{+}\pi^{+}$; $\omega$ mass resolution ($3.0$\%); the requirement 
on ${\mathcal R}$ ($3.0$\%) from $B^{-}\to D^{0}\pi^{-}$, $D^{0}\to K^{-}\pi^{+}$; MC statistics ($1.1$--$2.4$)\%, and the number of $B\overline{B}$ 
events in the data samples ($1.3$\%). For the ${\mathcal A}_{CP}$ measurement, the dominant systematic uncertainty comes from the fit parameters. 
Systematic uncertainties due to fit parameters are evaluated by varying each parameter by $\pm 1\sigma$ and adding in quadrature the resulting 
deviations from the central value. The total systematic uncertainty is the quadratic sum of all the above contributions. When varying each fit 
parameter by $\pm 1\sigma$ one at a time, the significance is recalculated and the lowest value is chosen as the final significance (shown in 
Table~\ref{prd06-2}) including the systematic uncertainty.\par          
While the branching fractions for $B^{+} \to\omega K^{+}$, $B^{0} \to \omega K^{0}$, and $B^{+} \to \omega \pi^{+}$ are of comparable size, 
that for $B^{0} \to \omega \pi^{0}$ is much smaller as compared to that for $B^{+} \to \omega \pi^{+}$. This is in agreement with both QCDF 
and PQCD pictures, which predict that only color-suppressed tree and penguin diagrams contribute to $B^{0} \to \omega \pi^{0}$~\cite{QCDF,PQCD,PQCD2}. 
We also determine the partial width ratio
\begin{eqnarray}
\frac{\Gamma (B^{+}\to \omega K^{+})}{\Gamma (B^{0}\to \omega K^{0})} & = & 1.7 \pm 0.3({\rm stat.})\pm 0.1({\rm sys.}),
\end{eqnarray} 
where the ratio of $B$ meson lifetimes $\frac{\tau_{B^{+}}}{\tau_{B^{0}}}$ is taken to be $1.076\pm 0.008$~\cite{PDG}. The systematic error of 
the partial width ratio is reduced because of the cancellation of several common systematic errors. No evidence for direct $CP$ violation 
is found for either $B^{+} \to \omega K^{+}$ or $B^{+} \to \omega \pi^{+}$. Our results on branching fractions and CP asymmetries are consistent 
with the recent measurements~\cite{babar3} from the BaBar collaboration.\par
We thank the KEKB group for the excellent operation of the accelerator, the KEK cryogenics group for the efficient operation of the solenoid, 
and the KEK computer group and the NII for valuable computing and Super-SINET network support. We acknowledge support from MEXT and JSPS (Japan); 
ARC and DEST (Australia); NSFC (contract No.~10175071 China); DST (India); the BK21 program of MOEHRD and the CHEP SRC program of KOSEF (Korea); 
KBN (contract No.~2P03B 01324, Poland); MIST (Russia); MHEST (Slovenia); SNSF (Switzerland); NSC and MOE (Taiwan); and DOE (USA). C.-M. Jen 
thanks Y.Y. Chang, H.Y. Cheng, C.H. Chen, C.K. Chua, and H.n. Li for invaluable instruction and discussions on particle physics phenomenology 
while at AS (Taiwan).\par

\begin{thebibliography}{99}
\bibitem{thrust} We define the thrust axis for a collection of particles as the axis that maximizes the sum of the magnitude of the 
longitudinal momenta with respect to the axis.
\end{thebibliography}
\end{document}